%% file: main.tex
\documentclass[runningheads]{llncs}

\usepackage[utf8]{inputenc}
\usepackage[english]{babel}
\usepackage{graphicx}
\usepackage{caption}
\usepackage{subcaption}
\usepackage{xcolor}
\usepackage{xurl}
\usepackage{url}
\usepackage{array}
\usepackage{amssymb}
\usepackage{pifont}
\usepackage{marvosym}
\usepackage{multirow}
\usepackage{booktabs}
\usepackage{enumitem}
\usepackage{hyperref}

\usepackage{xcolor}

\newcommand{\keywords}[1]{\par\addvspace\baselineskip
\noindent\keywordname\enspace\ignorespaces#1}

\begin{document}

\mainmatter

\title{A Decade of Climate Polarization on Brazilian YouTube using Language Models}
\titlerunning{A Decade of Climate Polarization on Brazilian YouTube}

% For anonymous submission, keep this block anonymized.
% Replace with author information in the camera-ready version.
\author{
Daniel Morais\inst{1} \and
Diego H. M. Magalh{\~a}es\inst{1} \and
Gabriel H. Silva\inst{1} \and
Andrea Failla\inst{3,4} \and
Val{\'e}ria de C. Santos\inst{2} \and
Helen C. S. C. Lima\inst{1} \and
Carlos H. G. Ferreira\inst{2}
}

\authorrunning{D. Morais et al.}

\institute{
Department of Computing and Systems,Universidade Federal de Ouro Preto,Brazil\\
\email{\{daniel.morais,diego.magalhaes,gabriel.hs\}@aluno.ufop.edu.br}
\and
Department of Computing,Universidade Federal de Ouro Preto,Brazil\\
\email{\{valeria.santos,helen,chgferreira\}@ufop.edu.br}
\and
ISTI, National Research Council (CNR),Italy\\
\email{andrea.failla@isti.cnr.it}
\and
Department of Computer Science, University of Pisa,Italy\\
}

\maketitle

\begin{abstract}
Online platforms have become arenas for the public contestation of climate change, shaping how scientific knowledge, denial, and uncertainty are expressed and disputed. Yet longitudinal evidence remains limited for YouTube, especially for Portuguese-language discourse. Addressing this gap, we characterize how climate stances are expressed and contested over time in a large corpus of Portuguese-language YouTube comments retrieved through Brazil-oriented climate-related searches. To support this analysis in a noisy, imbalanced, and low-resource setting, we collect more than 240{,}000 comments posted between 2014 and 2024 and formulate stance detection as a three-way classification task (\textit{Believer}, \textit{Denier}, and \textit{Inconclusive}). We operationalize stance attribution through a scalable self-training pipeline based on Llama~3.1, using Low-Rank Adaptation (LoRA) and hybrid instance selection to expand the training set with high-confidence pseudo-labeled examples while preserving class diversity. This approach improves coverage and class balance for minority and rhetorically complex classes, enabling large-scale stance attribution without extensive manual annotation. Our results show that polarization is marked by interactional asymmetries: denialist comments are less prevalent, but they are associated with a comparatively higher share of cross-stance contestation, while pro-consensus discourse is more strongly reinforced within stance-homogeneous threads. Qualitative analysis suggests that these dynamics are associated with distinct discursive frames, with pro-consensus discourse emphasizing governance and responsibility, while denialist discourse is rooted in institutional distrust and conspiratorial reasoning. Our study contributes as one of the first systematic characterizations of climate change stances in Portuguese-language YouTube comments and highlights how minority polarized discourse can become a focal point of conflict in a broader communicative environment.
%, largely inconclusive
\end{abstract}
%By combining scalable stance detection with longitudinal and interactional analysis

\input{sections/intro_pt}

\input{sections/related_work_pt}

\input{sections/meth_pt}
\input{sections/results_pt}

\input{sections/conclusion_pt}

% \begin{acks}
% TODO: preencher
% % 
% \end{acks}

\bibliographystyle{splncs04}
\bibliography{references}

\appendix
\section{Search Queries for Data Collection}\label{appendix:keywords}
\small
We use an extensive set of search queries to capture climate change discourse in Brazil. Because of the regional focus, most queries are in Portuguese, reflecting the platform's native discourse and reducing Anglophone bias. The strategy covers multiple dimensions of the debate, including the scientific basis of climate change, extreme events, human causes, policy and institutional references, and skepticism or denialist rhetoric. Terminological variation allows us to retrieve content associated with both scientific consensus and contestation, including references to official institutions (\textit{relatório IPCC resumo}\footnote{\url{https://www.ipcc.ch/}}, \textit{ONU mudanças climáticas}\footnote{\url{https://www.un.org/en/global-issues/climate-change}}) and misinformation or denialism (\textit{aquecimento global é farsa}, \textit{negacionismo climático}, \textit{dados manipulados aquecimento global}).

\paragraph{Queries.}
\textit{\footnotesize mudanças climáticas, aquecimento global, efeito estufa, crise climática, emer\-gência climática, clima extremo, recordes de temperatura, ondas de calor, ondas de frio e mudanças climáticas, chuvas intensas mudanças climáticas, secas e aquecimento global, enchentes mudanças climáticas, desastres naturais e clima, incêndios florestais mudanças climáticas, queimadas e aquecimento global, degelo das calotas polares, derretimento das geleiras, nível do mar subindo, aumento do nível do mar, impactos das mudanças climáticas, consequências do aquecimento global, causas das mudanças climáticas, mudanças climáticas causas humanas, mudanças climáticas causas naturais, atividade humana e clima, emissões de carbono, CO2 e aquecimento global, poluição e mudanças climáticas, indústria e aquecimento global, agronegócio e mudanças climáticas, desmatamento e mudanças climáticas, amazônia e aquecimento global, amazônia mudanças climáticas, clima no brasil, mudanças climáticas no brasil, clima extremo no brasil, chuvas no sul mudanças climáticas, seca no nordeste aquecimento global, mudanças climáticas são reais, aquecimento global é real, ciência das mudanças climáticas, consenso científico mudanças climáticas, IPCC mudanças climáticas, relatório IPCC resumo, cientistas falam sobre clima, mudanças climáticas explicadas, o que são mudanças climáticas, mudanças climáticas para iniciantes, mudanças climáticas é mentira, aquecimento global é farsa, mudanças climáticas conspiração, mudanças climáticas exagero, clima sempre mudou, sol causa aquecimento global, aquecimento global não existe, dados manipulados aquecimento global, agenda climática global, mudanças climáticas política, mudanças climáticas esquerda, mudanças climáticas direita, ambientalismo radical, imposto do carbono, agenda 2030 mudanças climáticas, ONU mudanças climáticas, acordo de paris, acordo de paris críticas, economia e mudanças climáticas, mudanças climáticas e energia, energia renovável e clima, carros elétricos aquecimento global, clima e capitalismo, negacionismo climático, fake news mudanças climáticas, mentiras sobre aquecimento global.}

\end{document}

%% file: sections/intro_pt.tex
\section{Introduction}

Online social media platforms play a central role in shaping public understanding of complex global crises \cite{malagoli2024twitter}. Rather than acting only as information repositories, platforms such as YouTube operate as interactional arenas where climate change is debated, contested, and frequently reframed or distorted~\cite{costa2025characterizing,Fonseca:2024}. These environments can disseminate scientific knowledge, but they can also amplify polarization and misinformation, affecting risk perception, trust in expertise, and the attribution of environmental responsibility~\cite{o2024certainty,aleix_bassolas_ec6b12ce}.

Prior work has documented polarization, denialist narratives, and emotional or conspiratorial framing in online climate discourse~\cite{Falkenberg2022Growing,Uyheng2021Mainstream,Beel2022Linguistic}. However, much of this evidence comes from Twitter/X and short observation windows, leaving less known about long-term dynamics and reply-based interactions on YouTube, where discussions are anchored in producer-generated videos and sustained through persistent comment threads that can foster reinforcement and conflict~\cite{Shapiro2014More,aleix_bassolas_ec6b12ce,VargasMeza2018Climate}.

Two related gaps motivate this study. First, despite Brazil’s central role in global climate debates and YouTube’s relevance as an information source in the country, we still know little about how climate-change stances are expressed over long periods in Portuguese-language YouTube discussions. Second, existing approaches often rely on manual qualitative analysis, sentiment-based measures, or short observation windows, which limits their ability to capture stance-based contestation in noisy, imbalanced, and reply-based comment environments. These gaps are especially important because YouTube discussions are not only collections of isolated comments, but persistent interactional spaces where agreement, denial, uncertainty, and contestation may unfold over time.

We address these gaps by characterizing the long-term evolution, interactional dynamics, and discursive framing of climate change stances in a Brazil-oriented corpus of Portuguese-language YouTube comments. We collect more than 240{,}000 comments posted between 2014 and 2024 on climate-related YouTube videos in Brazil, and combine scalable stance detection with quantitative and qualitative analyses to examine how belief, denial, and uncertainty toward climate science are expressed, reinforced, and contested in reply-based interactions over time. We guide the analysis through two research questions:
\\ \ \\
\textbf{RQ1} \textit{How are stances toward climate change distributed in Brazil-oriented YouTube discourse, and how do stance patterns and user interactions evolve over time?}
\\ \ \\
\textbf{RQ2} \textit{What discursive frames characterize polarized climate change stances in this corpus, and how do these frames contribute to conflict and polarization in reply-based interactions?}
\\ \ \\
To answer these questions, we formulate stance detection as a three-way classification task, distinguishing between \textit{Believer}, \textit{Denier}, and \textit{Inconclusive}. To scale the analysis, we develop a self-training pipeline based on Llama~3.1, fine-tuned with Low-Rank Adaptation (LoRA), and expand the training data through hybrid instance selection with high-confidence pseudo-labeled examples while preserving class diversity. We operationalize RQ1 by analyzing stance prevalence in the corpus and the yearly composition of replies to \textit{Believer}- and \textit{Denier}-initiated comments. We operationalize RQ2 by combining quantitative stance analysis with qualitative examination of representative polarized comment threads, focusing on recurring frames and their role in antagonistic and reinforcing reply exchanges.
\\ \ \\
Our results show that explicitly polarized stances constitute a substantial part of the debate and play an important role in structuring interaction. Denialist comments are less frequent, but they are associated with a larger share of cross-stance contestation, whereas pro-consensus discourse is more strongly reinforced within stance-homogeneous threads. Qualitative analysis further suggests that these dynamics are associated with distinct discursive frames: pro-consensus discourse emphasizes governance and responsibility, while denialist discourse is rooted in institutional distrust and conspiratorial reasoning. These findings contribute to social media and network analysis by linking large-scale stance detection, reply-based interaction dynamics, and discursive framing in an underexplored context.

%% file: sections/related_work_pt.tex
\section{Related Work}\label{sec:related_work}

Research on online climate discourse shows that social media platforms are key arenas for contesting scientific authority, political identity, and climate-related misinformation. Much of this literature focuses on Twitter/X, where studies document polarization around major climate events, the role of skeptical and right-wing actors, and the circulation of misinformation-oriented sources~\cite{Falkenberg2022Growing,Uyheng2021Mainstream}. Polarization is also reflected in language, affect, and stance: contentious climate debates exhibit distinctive linguistic markers~\cite{Beel2022Linguistic} and stance detection has been used to distinguish climate believers from skeptics or deniers~\cite{Upadhyaya2023MultiTask,perra2024quantifying}. Beyond Twitter/X, studies of Reddit and other communities show that climate discourse is organized around polarized subcommunities and recurring scientific, political, and conspiratorial frames~\cite{kathie_m__d_i__treen_c7ca2f43,or_elroy_1b9aafa1}. Together, these studies establish climate discourse as polarized and frame-dependent, but they remain concentrated on a limited set of platforms, mostly short-text environments, and relatively short observation windows.

This platform concentration matters because climate communication varies across online environments. YouTube is particularly relevant because producer-generated videos anchor persistent comment threads where scientific information, political commentary, and user debate coexist. Comparative work suggests that YouTube supports longer-form content and sustained discussion, whereas platforms such as TikTok tend to emphasize more emotional, self-referential, and action-oriented climate communication~\cite{Pera2024Shifting}. Earlier work also characterizes YouTube as a space for science communication in which users actively discuss and evaluate climate-related claims in comment sections~\cite{Shapiro2014More,VargasMeza2018Climate}. Yet large-scale and longitudinal analyses of climate stance on YouTube remain scarce, especially in Portuguese-language contexts and in the Global South. As a result, we know less about how climate stances evolve over time and how reply-based interactions structure reinforcement or conflict on this platform.

Addressing this gap requires scalable computational methods capable of handling noisy, informal, and imbalanced social media data. Recent NLP approaches have supported large-scale climate discourse analysis through deep learning pipelines, topic modeling, fine-grained stance classification, and LLM-assisted interpretation of latent themes~\cite{Cardoso2025Harnessing,Gokcimen2024Exploring,Vaid2022Towards,tunazzina_islam_66461054}. These methods demonstrate the value of computational approaches for studying climate discourse at scale, but they often rely on costly annotation, limited temporal scopes, or platforms dominated by short-text interactions.

Building on this literature, our work addresses two connected gaps. Empirically, we provide a decade-long characterization of climate-change stances in a Brazil-oriented Portuguese-language YouTube corpus, focusing on both stance prevalence and reply-based contestation. Methodologically and analytically, we combine scalable stance detection with an interaction-focused and frame-oriented analysis of how polarized positions are reinforced or contested in comment threads. This allows us to connect large-scale stance attribution to the interactional and discursive dynamics of climate polarization in an underexplored linguistic and regional context.

%% file: sections/meth_pt.tex
\section{Methodology}\label{sec:methodology}
% This section details the methodological framework adopted in this study, which is organized into four main stages. First, we describe the data collection and pre-processing procedures, outlining how YouTube videos metadata and their associated comments were retrieved and filtered. Second, we present the annotation guidelines used to curate a labeled dataset. Third, we introduce our self-training classification approach, leveraging the Llama~3.1 model. Finally, we analyze interaction patterns and the temporal dynamics of climate-related narratives. 

This section describes our four-stage methodology: data collection and pre-processing, manual annotation, self-training stance classification with Llama~3.1, and the analysis of temporal and interactional climate-discourse dynamics.

\subsection{Data Collection and Pre-processing}\label{sec:collection}

% Data collection was conducted using the YouTube Data API v3\footnote{\url{https://developers.google.com/youtube/v3}} to retrieve video metadata and associated comments. The temporal scope of the dataset spans from January~1,~2014 to December~31,~2024, covering a full decade of climate-related debate. To prioritize the Brazilian context, all search requests were restricted to the Brazilian region (\texttt{regionCode=BR}) and to Portuguese-language content whenever supported by the API, including Portuguese/Brazilian language parameters. No user profile, browser session, watch history, or other personalized information was provided in the collection process.

We collect video metadata and associated comments using the YouTube Data API v3\footnote{\url{https://developers.google.com/youtube/v3}}. The dataset spans January~1,~2014 to December~31,~2024, covering a decade of climate-related debate. To prioritize the Brazilian context, we restrict all searches to the Brazilian region (\texttt{regionCode=BR}) and, whenever supported by the API, to Portuguese-language content, including Portuguese/Brazilian language parameters. No user profile, browser session, watch history, or other personalized information is used.

% Videos were retrieved through keyword-based queries derived from prior literature, authoritative climate reports, and news, including terms such as ``climate change'', ``global warming'', ``greenhouse effect'', and ``climate hoax.'' All keywords and query configurations are documented in Appendix~\ref{appendix:keywords}. For each query and month in the observation window, we retrieved up to 50 videos returned by the API, following prior work~\cite{Costa:2025}. We used the API's default ordering criterion, \texttt{order=relevance}, which ranks resources according to their relevance to the search query. Since YouTube does not disclose the exact ranking formula, we do not treat this ordering as an externally reproducible or theory-driven measure of relevance. Instead, we use it as a platform-defined retrieval criterion, approximating the videos surfaced by YouTube's search infrastructure for climate-related queries under the specified temporal, regional, and language constraints. For each retained video, we collected the metadata made available by the API, including publication date, title, description, channel information, view count, like count, and comment count. For each associated comment, we retrieved the available textual content, publication timestamp, and parent--reply relationship.

We retrieve videos through keyword-based queries derived from prior literature, authoritative climate reports, and news, including terms such as ``climate change'', ``global warming'', ``greenhouse effect'', and ``climate hoax.'' All keywords and query configurations are documented in Appendix~\ref{appendix:keywords}. For each query and month in the observation window, we retrieve up to 50 videos returned by the API, following prior work~\cite{Costa:2025}. We use the API's default \texttt{order=relevance} criterion as a platform-defined retrieval mechanism, since YouTube does not disclose its ranking formula and the ordering cannot be treated as an externally reproducible or theory-driven relevance measure. Additionally, we did not adopt a user persona during the retrieval of videos in order to minimize potential biases introduced by personalized recommendations. For each retained video, we collect available metadata, including publication date, title, description, channel information, view count, like count, and comment count. For each comment, we collect its textual content, publication timestamp, and parent-reply relationship. This sampling strategy is designed to retrieve a large and thematically focused corpus of climate-related YouTube discussions in the Brazilian context.

% To ensure that the dataset reflected substantive discussions and genuine user interaction, we retained only videos containing more than 30 comments. This threshold excludes marginal or low-interaction videos and focuses the analysis on content that generated active debate. We then applied a content filtering pipeline to reduce semantic noise. First, we employed the \texttt{LangDetect} library\footnote{\url{https://pypi.org/project/langdetect/}} to retain only comments written in Portuguese. Subsequently, we conducted an exploratory data analysis to identify off-topic content retrieved through ambiguous keywords. Based on this inspection, we applied rule-based exclusions to remove videos unrelated to climate issues, such as electronic games or music videos that coincidentally contained terms associated with ``warming''.  After data collection and filtering, the resulting dataset comprises \emph{478} YouTube channels and \emph{1,020} videos, totaling \emph{247,514} comments written in Portuguese by \emph{137,585} distinct users.

To focus on substantive discussions, we retain only videos with more than 30 comments and apply a filtering pipeline to reduce semantic noise. We first use the \texttt{LangDetect} library\footnote{\url{https://pypi.org/project/langdetect/}} to detect and retain Portuguese comments. We then conduct exploratory analysis to identify off-topic content retrieved through ambiguous keywords and apply rule-based exclusions to remove videos unrelated to climate issues, such as games or music videos that coincidentally contain terms associated with ``warming''. After collection and filtering, the resulting dataset comprises 478 YouTube channels and 1,020 videos, totaling 247{,}514 Portuguese comments by 137{,}585 distinct users. A subset of these comments is reserved for manual annotation and model development, while the remaining comments form the inference corpus used for large-scale stance characterization.

Due to restrictions imposed by the YouTube API Services Developer Policies, which prohibit the redistribution of YouTube API data, the dataset cannot be released publicly. To support reproducibility, we will provide the list of video IDs, the complete query configuration, and the data collection scripts, allowing authorized researchers to rehydrate the dataset directly from the official YouTube Data API v3.\footnote{\url{https://developers.google.com/youtube/terms/developer-policies}}

%\CH{Daniel. Você tem o modelo aí? Consegue fazer algo similar a isso: https://huggingface.co/gseovana/llama-vaccine-stance-ptbr-lora}

\subsection{Comment Annotation}\label{sec:annotation}

% With the dataset defined, the next step consisted of manually annotating a subset of comments, a critical stage for both the development and validation of the stance detection model. The objective was to infer explicit comment-level positions regarding climate change by classifying each comment into one of three categories: \emph{Believer}, \emph{Denier}, or \emph{Inconclusive}.

% To construct the labeled set, we randomly sampled 4{,}200 candidate comments from the pool of eligible comments after pre-processing. During quality control, 123 instances were removed because they were duplicate, invalid, or empty after text normalization, resulting in 4{,}077 comments submitted to annotation. The annotation unit was the individual comment, including both top-level comments and replies. We defined clear and mutually exclusive annotation guidelines, assigning each comment to one of the following categories:

We manually annotate a subset of comments to develop and validate the stance detection model. The goal is to infer explicit comment-level positions toward climate change by assigning each comment to one of three categories: \emph{Believer}, \emph{Denier}, or \emph{Inconclusive}. From the pool of eligible comments after pre-processing, we randomly sample 4{,}200 candidates. Quality control removes 123 duplicate, invalid, or empty instances after text normalization, resulting in 4{,}077 comments submitted to annotation. The annotation unit is the individual comment, including both top-level comments and replies. We define mutually exclusive guidelines as follows:

\begin{itemize}[leftmargin=0.30cm]
\small
    \item \textbf{Believer:} Comments that acknowledge climate change and agree with the scientific consensus, including environmental concern, defense of scientific evidence, or criticism of harmful practices such as deforestation or wildfires.\looseness=-1
    \item \textbf{Denier:} Comments that express skepticism, deny climate change, downplay its impacts, or reject anthropogenic responsibility, including claims that global warming is a ``hoax'', a ``lie'', a natural cycle, or a conspiracy.\looseness=-1
    \item \textbf{Inconclusive:} Comments whose stance cannot be determined from the text alone, including ambiguous statements, generic agreement or disagreement without a clear target, irony without sufficient cues, and tangential discussions (e.g., partisan disputes) where climate change is not central.\looseness=-1
\end{itemize}

To match the model input, annotation relies only on comment text. Annotators do not access the parent video, title, channel metadata, parent comment, or surrounding thread; consequently, replies are not labeled according to the stance of the parent comment. Generic agreement or disagreement (e.g., ``I agree'', ``exactly'', or ``this is a lie''), irony, and sarcasm are labeled as \emph{Inconclusive} whenever the target stance toward climate change cannot be inferred from the comment alone. This conservative protocol may underestimate context-dependent stance expression, but avoids labels based on information unavailable to the classifier.

% The annotation process was conducted independently by two authors, after a preliminary calibration round to align the interpretation of the guidelines. Inter-annotator agreement was evaluated using Cohen's Kappa coefficient, yielding a value of 0.88, which indicates a strong level of agreement according to established benchmarks in the literature~\cite{McHugh:2021}. Disagreements were adjudicated by a third author. In cases of persistent disagreement, where consensus could not be reached even after adjudication, the corresponding instances were excluded from the dataset, totaling 69 comments.

Two authors independently annotate the comments after a calibration round to align the guidelines. Inter-annotator agreement is measured with Cohen's Kappa, yielding 0.88, which indicates strong agreement~\cite{McHugh:2021}. Disagreements are adjudicated by a third author; 69 comments for which consensus cannot be reached are excluded. The final labeled dataset comprises 4{,}008 comments and is substantially imbalanced: \emph{Inconclusive} accounts for 2{,}241 comments (55.9\%), followed by \emph{Believer} with 1{,}097 (27.4\%) and \emph{Denier} with 670 (16.7\%).

\subsection{Model Training and Evaluation}\label{sec:model}

We use the manually annotated dataset from Section~\ref{sec:annotation} to train an initial supervised stance classifier based on Llama~3.1~8B~\cite{Dubey2024}. We choose this open-access model because LLMs perform well in semantic understanding and text classification task, especially in noisy social media discourse~\cite{Costa:2025}, while also supporting reproducible academic research. To adapt it to stance detection, we add a shallow task-specific classification head on top of the frozen Llama backbone: a single linear layer maps the final hidden representation to an $n$-dimensional logits vector, one dimension per stance class, followed by a softmax.

To reduce computational cost, we fine-tune the model with Low-Rank Adaptation (LoRA), which adds trainable low-rank matrices while keeping the base weights frozen~\cite{Hu:2022}. We configure LoRA with rank $r=16$, scaling factor $\alpha=32$, and dropout $0.1$, inserting the adapters only into the attention projection matrices (\textit{query, key, value, and output}). We further combine LoRA with \textit{Quantized LoRA (QLoRA)}~\cite{Dettmers:2023}, quantizing the base model to 4 bits with the \textit{BitsAndBytes} library, the \textit{NF4} scheme, and double quantization. This setup reduces memory usage and enables fine-tuning Llama~3.1~8B on a single commodity GPU.

Training uses weighted cross-entropy, with class weights inversely proportional to class frequencies, to mitigate imbalance and reduce majority-class bias~\cite{Kamath:2024}. We also apply label smoothing to discourage overconfident predictions. Optimization uses AdamW with an initial learning rate of $2 \times 10^{-4}$, a cosine learning rate scheduler, and a warmup phase~\cite{Costa:2025}. We train for up to 50 epochs and apply early stopping based on validation loss, stopping after 5 epochs without improvement.

We evaluate performance with stratified 5-fold cross-validation. The supervised baseline is evaluated on the manually annotated dataset, whereas the enhanced model, after the self-training procedure described below, is evaluated with the same protocol on the expanded dataset. Macro F1-score is the primary metric because it gives equal weight to all classes under imbalance. We also report accuracy and per-class precision and recall.

\subsubsection{Self-Training and Hybrid Instance Selection}

Given the limited size and strong imbalance of the manually annotated dataset, we use self-training as a weak-supervision strategy to improve coverage before large-scale corpus annotation. Our goal in this stage is not to create a new manually validated test set, but to construct a more robust final classifier for assigning stance labels to a fixed historical corpus of Brazilian YouTube comments. This design follows the general self-training principle of using a supervised model to assign labels to unlabeled instances, while controlling the inclusion of pseudo-labeled examples to reduce noise and avoid reinforcing early model biases~\cite{Amini:2025}.

This choice is motivated by the low-resource setting, since manually labeled examples are scarce, retaining part of them outside the final training stage would reduce the amount of human supervision available for the model that is ultimately used to label the complete corpus. The manually annotated data therefore serve as the anchor for the stance definitions, while pseudo-labeled instances are used only as additional weak supervision. The initial classifier is then applied to previously unlabeled comments. For each comment, the model outputs a probability distribution over the three stance classes, and the pseudo-label corresponds to the class with the highest predicted probability \cite{Oliveira:2026}. 

To reduce the risk of propagating noisy labels, we retain only predictions whose confidence exceeds a class-wise threshold of $0.75$. Applying the threshold separately within each predicted class helps prevent the expanded training set from being dominated by the majority class or by the easiest-to-classify instances. Confidence-based filtering alone, however, may reinforce early model biases by selecting highly redundant examples and concentrating the pseudo-labeled set in narrow regions of the unlabeled data distribution. To mitigate this issue, we combine confidence filtering with controlled random sampling within the eligible pool. Thus, confidence filtering acts as a noise-control mechanism, whereas random sampling functions as a diversity-preserving mechanism. This hybrid design is consistent with prior self-training literature, which emphasizes both the need to limit incorrect pseudo-labels and the risk of selection bias in overly deterministic pseudo-labeling strategies~\cite{Amini:2025}.

We further sample pseudo-labeled instances in inverse proportion to the original class distribution, increasing the relative presence of minority stance categories. This is particularly important because \emph{Believer} and \emph{Denier} comments are substantively central to the downstream analysis but underrepresented in the manual dataset. For training set expansion, we incorporate pseudo-labeled instances equivalent to an additional 50\% of the manually labeled dataset. The final expanded training set consists of 1{,}738 instances labeled as \emph{Believer}, 1{,}719 as \emph{Denier}, and 2{,}555 as \emph{Inconclusive}. The resulting model should therefore be understood as a self-trained classifier for large-scale stance attribution rather than as a purely supervised model evaluated on an independent gold-standard test set. Pseudo-labels are not treated as human annotations. Conversely, they are used as weak labels to improve the final model’s coverage and class balance before applying it to the full YouTube corpus. The resulting self-trained LoRA adapter is publicly available on Hugging Face.\footnote{\url{https://huggingface.co/danielangelo1/llama-climate-change-stance-ptbr-lora}} All experiments were conducted on a server equipped with an Intel Xeon Gold 6442Y CPU at 2.6 GHz, an NVIDIA A40 GPU with 48 GB of VRAM, and 512 GB of RAM.

\subsection{Large-Scale Stance Characterization}
\label{sec:characterization}

After obtaining the final self-trained model, we apply it in inference mode to the remaining previously unlabeled comments, excluding the subset used for manual annotation and model development. Each comment receives one stance label (\textit{Believer}, \textit{Denier}, or \textit{Inconclusive}) according to the highest predicted probability, producing a consistently labeled corpus. We use this corpus to characterize climate discourse at scale by examining stance distribution, reply-based interaction patterns, and discursive framing over the ten-year period.

For RQ1, we quantify class prevalence in the inference corpus and examine the
yearly composition of replies to \textit{Believer}- and \textit{Denier}-initiated comments,
capturing reinforcement and cross-stance contestation over time. For RQ2, we identify high-engagement polarized contexts and examine representative parent-reply exchanges, focusing on recurring interpretive patterns through which commenters define climate change, attribute responsibility, and contest epistemic authority.

%% file: sections/results_pt.tex
\section{Results}\label{sec:resultados}
We organize the findings into three parts: classifier performance, stance and interaction dynamics, and qualitative patterns in polarized narratives.

\subsection{Classifier Diagnostics}\label{sec:model_performance}

We evaluate the proposed approach using stratified 5-fold cross-validation, comparing a supervised baseline with the \textit{enhanced model} obtained through self-training and hybrid instance selection. We report accuracy, macro-averaged metrics, and per-class precision, recall, and F1-score. All values correspond to means across folds with 95\% confidence intervals.

Table~\ref{tab:model_comparison_detailed} summarizes the results. We assess statistical significance with a t-test ($\alpha=0.05$), which is appropriate for small-sample comparisons and does not assume equal variances~\cite{Welch:1947}. Statistically significant differences are marked with an asterisk (*).

\begin{table}[t]
\centering
\footnotesize
\setlength{\tabcolsep}{5pt}
\renewcommand{\arraystretch}{0.92}
\caption{Comparison between baseline and enhanced models under 5-fold cross-validation. Values are mean $\pm$ 95\% CI. Statistically significant differences under t-test ($p<0.05$) are marked with *.}
\label{tab:model_comparison_detailed}
\begin{tabular}{lcc}
\toprule
\textbf{Metric} & \textbf{Baseline} & \textbf{Enhanced} \\
\midrule
Accuracy* & $0.719 \pm 0.025$ & $\mathbf{0.795 \pm 0.019}$ \\
Macro-Precision* & $0.676 \pm 0.026$ & $\mathbf{0.792 \pm 0.026}$ \\
Macro-Recall* & $0.687 \pm 0.032$ & $\mathbf{0.791 \pm 0.014}$ \\
Macro-F1* & $0.677 \pm 0.029$ & $\mathbf{0.789 \pm 0.020}$ \\
\midrule
Believer Precision* & $0.619 \pm 0.095$ & $\mathbf{0.756 \pm 0.048}$ \\
Believer Recall* & $0.614 \pm 0.087$ & $\mathbf{0.705 \pm 0.042}$ \\
Believer F1* & $0.610 \pm 0.026$ & $\mathbf{0.729 \pm 0.024}$ \\
\midrule
Denier Precision* & $0.575 \pm 0.060$ & $\mathbf{0.783 \pm 0.073}$ \\
Denier Recall* & $0.658 \pm 0.103$ & $\mathbf{0.845 \pm 0.064}$ \\
Denier F1* & $0.611 \pm 0.060$ & $\mathbf{0.810 \pm 0.026}$ \\
\midrule
Inconclusive Precision & $0.833 \pm 0.027$ & $0.837 \pm 0.034$ \\
Inconclusive Recall & $0.788 \pm 0.057$ & $0.823 \pm 0.063$ \\
Inconclusive F1 & $0.809 \pm 0.017$ & $0.828 \pm 0.016$ \\
\bottomrule
\end{tabular}
\vspace{-0.2cm}
\end{table}

At the macro level, the enhanced model shows higher diagnostic scores across all evaluated metrics. Macro-F1 increases from $0.677 \pm 0.029$ to $0.789 \pm 0.020$ ($p<0.001$), an absolute gain of 0.112. Similar gains in Macro-Precision and Macro-Recall indicate that the improvement is not driven by a single metric, but reflects a more balanced error profile. The narrower confidence intervals, especially for Macro-Recall, also suggest greater stability across folds.

At the class level, gains for \textit{Believer} and \textit{Denier} are statistically significant across precision, recall, and F1-score. This result is important because these are the minority and substantively central classes for the downstream analysis. In contrast, improvements for \textit{Inconclusive} are positive but not statistically significant, indicating that the proposed strategy mainly improves minority-class discrimination without materially changing performance on the already well-classified majority class.

The class-level breakdown also clarifies why the enhanced model is preferable for large-scale stance attribution. In the baseline model, low \textit{Believer} recall ($0.614 \pm 0.087$) would underestimate pro-consensus participation, while low \textit{Denier} precision ($0.575 \pm 0.060$) would inflate denial estimates and potentially overstate polarization. The enhanced model mitigates both risks: \textit{Believer} F1 increases from $0.610 \pm 0.026$ to $0.729 \pm 0.024$, and \textit{Denier} F1 increases from $0.611 \pm 0.060$ to $0.810 \pm 0.026$. Overall, these diagnostics suggest that the hybrid self-training strategy yields a more balanced classifier for large-scale stance attribution, especially for minority and rhetorically complex classes.

\subsection{Large-Scale Stance Distribution and Interaction Dynamics}
\label{sec:stance_distribution}

To address RQ1, we apply the final enhanced model to the inference corpus,
which comprises 242{,}367 previously unlabeled comments after excluding the subset reserved for manual annotation and model development. The distribution is dominated by \textit{Inconclusive} comments, with 157{,}414 instances (64.9\%), followed by \textit{Believer} with 46{,}867 (19.3\%) and \textit{Denier} with 38{,}086 (15.7\%).

The predominance of \textit{Inconclusive} comments suggests that discussions on Brazilian YouTube frequently intersect with political disputes, identity signaling, irony, contextual remarks, and other forms of commentary that do not explicitly align with either acceptance or rejection of the scientific consensus. 
Nevertheless, 84{,}953 comments, roughly one third of the corpus, express a clear stance toward climate change. This indicates that explicitly positioned discourse constitutes a substantial component of the debate, even within a broader communicative environment where many comments remain ambiguous, contextual, or tangential.

These results suggest that, within the retrieved corpus, polarization is not only a matter of stance prevalence but also of how stance categories are embedded in reply-based interaction. Climate change is often discussed alongside broader social, political, and cultural issues, while still sustaining a sizable core of explicitly polarized participation. We therefore examine whether these stance groups also differ in how they structure interaction within comment threads.

\begin{figure}[t]
    \centering
    \includegraphics[width=0.8\linewidth]{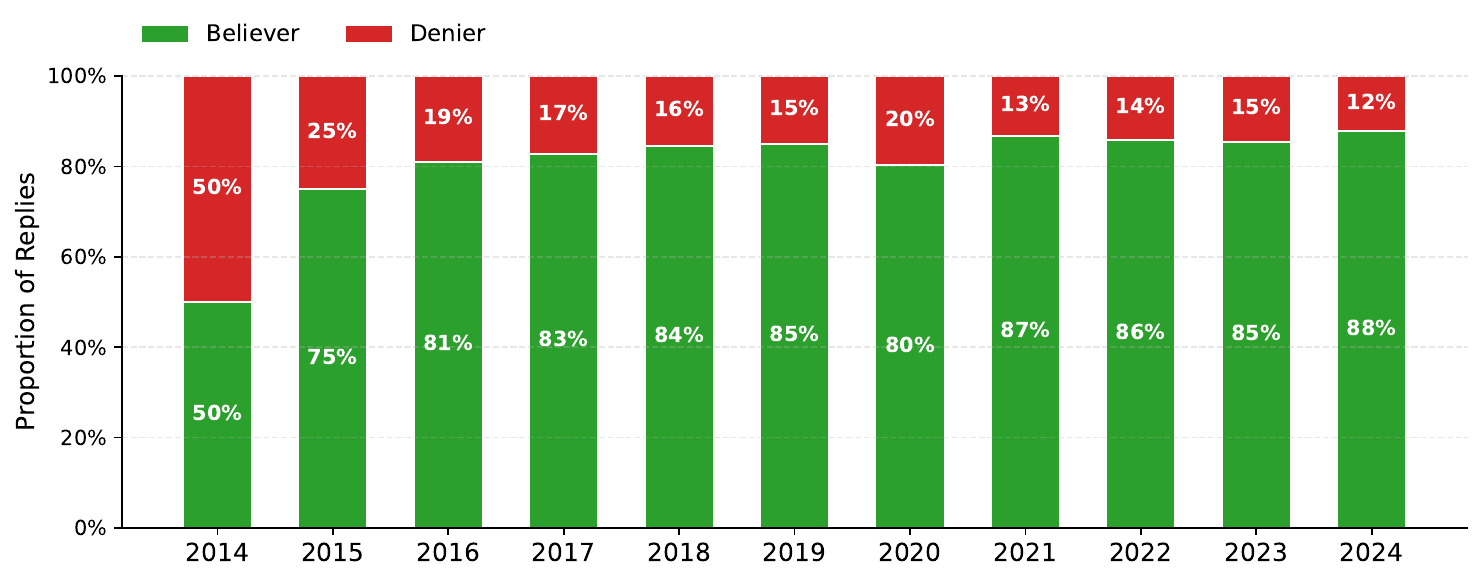}
    \caption{Temporal evolution of reply stance proportions within threads initiated by \textit{Believer} parent comments (2014--2024).}
    \label{fig:believer_threads}
    \vspace{-0.3cm}
\end{figure}

\begin{figure}[t]
    \centering
    \includegraphics[width=0.8\linewidth]{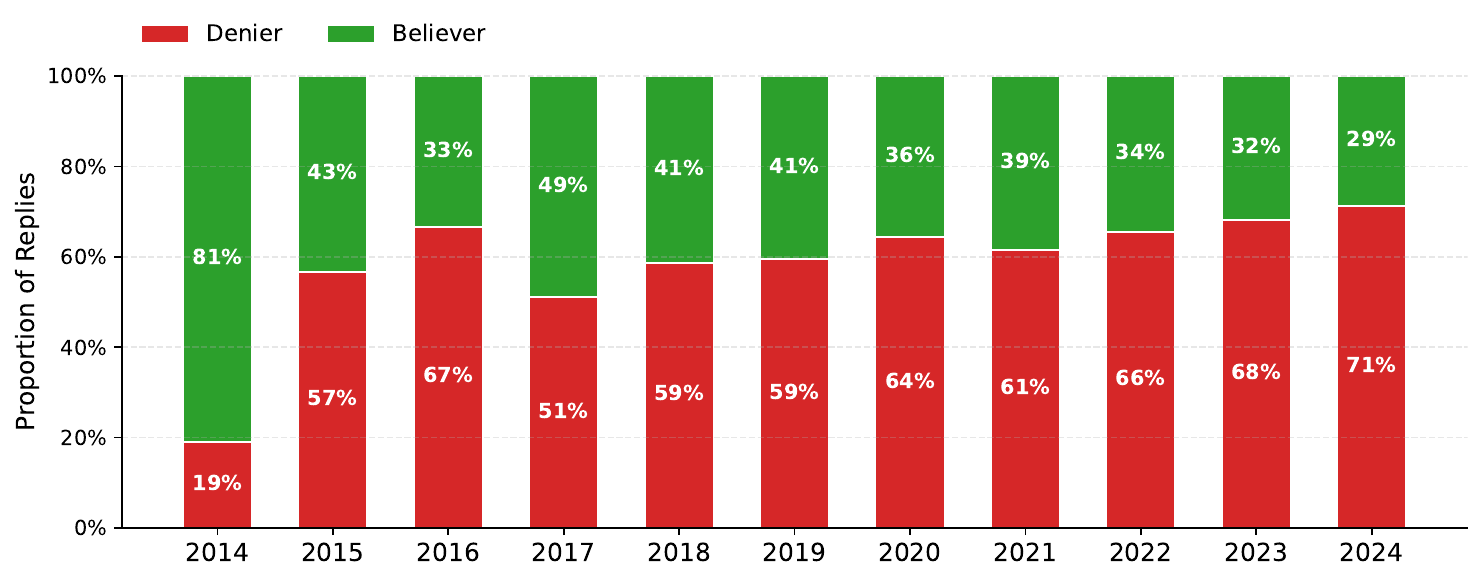}
    \caption{Temporal evolution of reply stance proportions within threads initiated by \textit{Denier} parent comments (2014--2024).}
    \label{fig:denier_threads}
    \vspace{-0.6cm}
\end{figure}

We analyze the yearly composition of replies to \textit{Believer}- and \textit{Denier}-initiated comments, focusing on explicitly polarized threads. Figure~\ref{fig:believer_threads} shows that threads initiated by \textit{Believer} comments are consistently dominated by \textit{Believer} replies. From 2016 onward, about 80\% to 88\% of replies in these threads come from the same stance, while \textit{Denier} replies generally remain below 20\%. This pattern indicates strong within-stance reinforcement and limited cross-stance engagement. Figure~\ref{fig:denier_threads} shows a more contested pattern. Although \textit{Denier} replies form the majority in most years, their share is lower and more variable, while \textit{Believer} replies account for a substantial fraction of the discussion. From 2015 onward, \textit{Believer} replies typically represent about 29\% to 49\% of replies in \textit{Denier}-initiated threads, revealing sustained cross-stance engagement.

Taken together, these figures show a stable interactional asymmetry within the retrieved corpus. \textit{Believer} comments mainly generate within-group reinforcement, whereas Denier comments more often attract opposition. Although denialist comments are less prevalent overall, they attract a comparatively higher share of cross-stance contestation.

\subsection{Discursive Frames and High-Engagement Polarization}
\label{sec:narrative_frames_lightning_rods}

To address RQ2, we examine the discursive content underlying the interactional asymmetries identified above. Using the stance labels produced by the enhanced model, we analyze how climate change is framed, justified, and contested across polarized positions, especially in high-engagement contexts. We use \textit{discursive frame} to refer to a recurring interpretive pattern through which commenters define the climate issue, attribute responsibility, evaluate credible authorities, and assign political meaning to climate change.

Within the \textit{Believer} class, discourse rarely centers on explicitly defending climate science itself. Instead, recurrent terms such as ``government,'' ``agribusiness,'' ``deforestation,'' and ``nature'' frame climate change as a problem of public governance, economic responsibility, and environmental justice. In this frame, scientific consensus is largely treated as a premise for broader claims about political inaction, predatory economic practices, and collective accountability. In contrast, \textit{Denier} rhetoric follows a conspiratorial and anti-systemic logic. Terms such as ``globalism,'' ``HAARP,'' ``hoax,'' ``control,'' and ``lie'' suggest that rejection of climate science is tied to distrust of global institutions, expert authority, and political elites. Rather than engaging primarily with empirical evidence, denialist discourse often shifts the debate toward intent, power, and legitimacy, reframing climate science as an instrument of ideological manipulation or social control.

To examine how these frames become amplified, we focus on high-engagement videos in the top comment-volume decile for both \textit{Believer} and \textit{Denier} discussions. Highly engaged \textit{Believer} videos tend to frame climate change through governance failure, economic risk, territorial vulnerability, and scientific authority. Examples include videos linking Amazon exploitation and environmental destruction to global warming, agricultural risk, desertification, and Brazil's disproportionate exposure to climate impacts. By contrast, highly engaged \textit{Denier} videos often contest anthropogenic causality, invoke natural or cosmic explanations, or present climate science as politically motivated. Examples include videos associating climate change with magnetic pole reversal, HAARP, globalist agendas, or broader narratives of institutional control. Notably, even videos presenting concrete visual evidence of climate impacts can become discursive battlegrounds, attracting denialist scrutiny rather than producing consensus.

\begin{table*}[t]
\scriptsize
\centering
\setlength{\tabcolsep}{4pt}
\renewcommand{\arraystretch}{0.95}
\caption{Examples of parent comments attracting antagonistic engagement from the opposing stance. The symbol \# indicates the total number of replies from the antagonistic class.}
\label{tab:conflict_examples}
\begin{tabular}{p{0.08\linewidth} | p{0.39\linewidth} | p{0.42\linewidth} | c}
\toprule
\textbf{Parent Stance} & \textbf{Parent Comment (translated)} & \textbf{Example Antagonistic Reply (translated)} & \textbf{\#} \\
\midrule
\textit{Believer} &
``For at least 20 years, environmentalists and scientists have been warning us. But human greed and deforestation have not decreased. Nature has reached its limit.'' &
``Don’t fall for this. These are natural planetary cycles. Human activity did not cause all of this.'' &
11 \\
\midrule
\textit{Believer} &
``I can’t understand how anyone still doubts global warming. Some people insist on spreading misinformation. Thankfully, we have channels like Nerdologia and Pirula doing science communication.'' &
``Look up Climategate and you’ll understand why many people started distrusting the IPCC. Billions go into NGO funding. It has become an industry.'' &
11 \\
\midrule
\textit{Denier} &
``The problem with these studies is bias. Governments usually use this for politics. [...] It is more realistic for us to adapt to the new climate than to expect governments to truly solve the problem.'' &
``Aviation accounts for only 3\%. You use the phrase ‘doing politics’ without even understanding what it means. A sad reality of a population that was not properly educated.'' &
44 \\
\midrule
\textit{Denier} &
``Have a debate with Prof. Ricardo Felício. If 97\% believe everything about global warming, it should be easy for you to refute all of his arguments.'' &
``There are hundreds of professors in Brazil who are specialists in the field and say exactly the opposite. If you want to use a PhD’s authority as an argument, then you lose. Global warming is a consensus.'' &
27 \\
\bottomrule
\end{tabular}
\vspace{-0.7cm}
\end{table*}

Table~\ref{tab:conflict_examples} illustrates how specific parent comments operate as triggers for concentrated antagonistic engagement\footnote{All examples were translated from Portuguese and paraphrased for privacy-preserving presentation, while retaining their original meaning, stance, and argumentative structure.}. Although these exchanges represent a minority of the overall discourse, they reveal how cross-stance contestation clusters around highly visible narrative cues. Taken together, the results suggest that polarization in Brazilian YouTube climate discourse involves not only disagreement over evidence, but also competing epistemic and political frames: \textit{Believers} mobilize scientific authority to demand responsibility and collective action, whereas \textit{Deniers} frame climate science as an instrument of ideological or institutional control. Thus, a relatively small subset of polarized discourse can structure conflict and visibility within a much larger and predominantly inconclusive communicative landscape.

%% file: sections/conclusion_pt.tex
\section{Conclusion and Future Work}\label{sec:conclusao}
This study characterizes a decade of climate discourse in a Brazil-oriented corpus of Portuguese-language YouTube comments. Regarding RQ1, polarization reflects both stance prevalence and interactional asymmetry. Although most comments are classified as \textit{Inconclusive}, polarized stances disproportionately structure debate. Denialist comments are less prevalent but attract more cross-stance contestation, making denial a recurring catalyst of conflict.

Regarding RQ2, polarized stances rely on distinct discursive frames. Pro-consensus discourse emphasizes governance, responsibility, and collective action, whereas denialist discourse reflects institutional distrust, conspiratorial reasoning, and suspicion toward expert authority. This contrast suggests that countering denial requires more than scientific information, combining evidence with trust-building, institutional transparency, and context-sensitive engagement. Methodologically, our self-training approach enables scalable stance detection in noisy, imbalanced, and low-resource settings, supporting analyses that would be difficult through manual annotation alone.

Future work should compare Global South countries to assess whether similar asymmetries emerge under different political and environmental conditions. We also aim to apply narrative detection techniques to track evolving narratives and support deeper qualitative analysis. Finally, linking online discourse to environmental disasters, elections, policy debates, and media coverage could clarify how climate polarization responds to heightened public salience.